\documentclass[twocolumn,twoside]{IEEEtran} 

\usepackage{amsmath,amssymb,amsfonts}
\usepackage[final]{graphicx}
\usepackage{psfrag}
\usepackage[numbers,sort&compress]{natbib}
\usepackage{flushend}
\usepackage{epstopdf}
\usepackage{multirow}
\usepackage{color}
\usepackage{booktabs}

\makeatletter
\def\blfootnote{\xdef\@thefnmark{}\@footnotetext}
\makeatother

\begin{document}
\title{\Huge{On the Equivalence between Interference and \\ Eavesdropping in Wireless Communications}}
\author{G. Gomez, F. J. Lopez-Martinez, D. Morales-Jimenez, and M. R. McKay}
\maketitle
\begin{abstract}
We show that the problem of analyzing the outage probability in cellular systems affected by co-channel interference and background noise is mathematically equivalent to the problem of analyzing the wireless information-theoretic security in terms of the secrecy outage probability in fading channels. Hence, these both apparently unrelated problems can be addressed by using a common approach. We illustrate the applicability of the connection unveiled herein to provide new results for the secrecy outage probability in different scenarios.

\end{abstract}


\section{Introduction}
\blfootnote{Copyright (c) 2014 IEEE. Personal use of this material is permitted. However, permission to use this material for any other purposes must be obtained from the IEEE by sending a request to pubs-permissions@ieee.org.}
\blfootnote{This work has been partially supported by Junta de Andalucia under project P11-TIC-7109, the Spanish Government and FEDER under projects TEC2013-44442-P and COFUND2013-40259, the University of Malaga and the European Union under Marie-Curie COFUND U-mobility program (ref. 246550). The work of D. Morales-Jimenez and M. R. McKay was supported by the Hong Kong Research Grants Council under grant number 616713.
\\ \indent G. Gomez and F.J. Lopez-Martinez are with Dpto. Ingenieria de Comunicaciones, University of Malaga, 29071 Malaga, Spain. (email \{ggomez,fjlopezm\}@ic.uma.es). D. Morales-Jimenez and M. R. McKay are with Dept. Electronic and Computer Engineering, Hong Kong University of Science and Technology, Clear Water Bay, Kowloon, Hong Kong. (email \{eedmorales, eemckay\}@ust.hk)
}

The characterization of the performance of wireless communication systems in the presence of co-channel interference (CCI) has been a matter of intense research for more than 20 years \cite{Abu1991}, due to the advent of digital cellular communication standards whose performance is limited by the effect of such interference. Many authors have dealt with the effects of CCI in a plethora of scenarios, considering different numbers of interfering signals that can be independent or correlated, terminals equipped with one or more receive antennas, as well as considering diverse families of distributions to characterize the fading in the desired and the interfering links, in the presence or absence of background noise (BN). Some examples can be found in classical references in communication theory \cite{Tellambura1999,Tellambura1999b,Annamalai2001,Kang2004}, as well as in more contemporary works \cite{Romero2007,Romero2008,Morales2012,Paris2013,Ermolova2014}.

In a different context, an apparently unrelated problem is the characterization of the secure communication between two legitimate peers (usually referred to as Alice and Bob) through a wireless link in the presence of an eavesdropper (referred to as Eve), that observes this communication through a different link. In contrast to what is known for the Gaussian wiretap channel \cite{Leung1978}, fading allows for a secure communication between Alice and Bob even in the case when the SNR observed at Eve is larger than the SNR at the legimitate receiver \cite{Barros2006,Bloch2008}. In the last few years, many researchers have worked in the characterization of the physical layer security in this scenario, using the secrecy outage probability and the probability of strictly positive secrecy capacity as performance metrics. Analytical results are available for some of the most usual fading distributions: Rayleigh \cite{Bloch2008}, Nakagami-$m$ \cite{Sarkar2009}, Rician \cite{Liu2013}, Hoyt \cite{Romero2014}, lognormal \cite{Liu2013b} or two-wave with diffuse power \cite{Wang2014} fading models, where the effect of having additional antennas at the eavesdropper has been also investigated.

\begin{figure}[t]
\begin{center}
\includegraphics[width=.9\columnwidth]{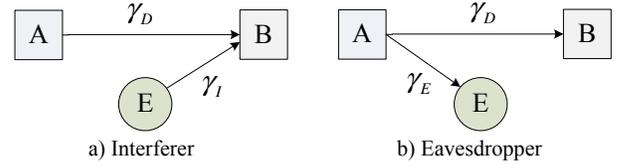}
\caption{A typical wireless communication scenario in the presence of: a) an interferer, b) an eavesdropper.}
\label{fig1}
\end{center}
\end{figure}

The first of the aforementioned problems can be seen as a transmitter $A$ willing to communicate with the receiver $B$ in the presence of an external interferer $E$, which accounts for the effect of aggregate CCI. Hence, we observe \textit{two transmitters} and \textit{one receiver}, as depicted in \mbox{Fig. \ref{fig1}a)}. Conversely, in the second problem we see a transmitter $A$ willing to communicate with the receiver $B$ in the presence of an external observer $E$. Therefore, we can identify \textit{one transmitter} and \textit{two receivers}, as shown in Fig. \ref{fig1}b).

Motivated by this duality, we wonder whether there is any connection between both scenarios that facilitates the analysis of the latter using the extensive knowledge available for the former. In this paper, we show that \textit{both problems are in fact mathematically equivalent}; this relevant connection has not been reported in the literature before, to the best of our knowledge.

We also show that the duality between both scenarios holds not only when using the classical definition of secrecy capacity \cite{Barros2006} but also with the alternative formulation in \cite{Zhou2011}, which is more suitable for adaptive scenarios in which channel state information (CSI) of Bob is available at the transmitter. From this duality, we show that the calculation of the outage probability (OP) with CCI and BN in the presence of $L_e$ interferers is equivalent to the calculation of the secrecy capacity OP in the presence of an eavesdropper equipped with $L_e$ antennas and performing maximal ratio combining (MRC), up to a simple scaling of the underlying random variables. Similarly, the calculation of the OP in interference-limited scenarios (i.e. neglecting the BN) is equivalent to the problem of computing the probability of strictly positive secrecy capacity.

\section{System Model I: CCI and BN}

\label{Scenario I}
Let us first consider the case where all the agents are single antenna transceivers, and there is only one interfering signal. We characterize the communication links in Fig. \ref{fig1}a) in terms of two instantaneous power ratios at the receiver side (B), the SNR ${\gamma_d=P_T\left|h_d\right|^2/N_0}$ and the interference-to-noise ratio (INR) ${\gamma_i=P_I\left|h_{i}\right|^2/N_0}$, where $P_T$ and $P_I$ indicate the transmission power of A and the interferer I, respectively, $N_0$ denotes the noise power at B, whilst $h_d$ and $h_i$ represent the complex fading channel gain for the desired and interfering links respectively, which incorporate the effects of attenuation due to path loss (i.e. proportional to $d^{-\alpha}$, where $d$ is the distance between transmitter and receiver, and $\alpha$ is the path loss exponent \cite{Barros2006}).

In the most general scenario with both CCI and BN, the usual performance metric is the OP of the signal-to-interference-plus-noise ratio (SINR), defined as the probability of the instantaneous SINR falling below a given threshold $\gamma$, i.e.: 
\begin{equation}
\label{eq1}
OP_{NI} (\gamma) = \Pr \left\{ {\frac{{P_T\left|h_d\right|^2 }}{{P_I\left|h_i\right|^2 {\rm{ + }}N_0}}{\rm{ < }}\gamma } \right\}  = \Pr \left\{ {{{\gamma_d }}{\rm{ < }}\gamma {{(\gamma_i {\rm{ + }}1)}}} \right\} .
\end{equation}
In case of interference-limited scenarios (i.e. $N_0=0$), the previous expression reduces to
\begin{equation}
\label{eq2}
OP_{I} (\gamma) = \Pr \left\{ {\frac{{P_T\left|h_d\right|^2 }}{{P_I\left|h_i\right|^2 }}{\rm{ < }}\gamma } \right\} = \Pr \left\{ {{{\gamma_d }}{\rm{ < }}(\gamma \cdot {{\gamma_i}}}) \right\}.
\end{equation}
Note that since we are neglecting the BN in (\ref{eq2}), ${{P_T\left|h_d\right|^2 }}/{{P_I\left|h_i\right|^2 }}$ represents now a signal-to-interference ratio (SIR). For simplicity we still use the same nomenclature as in (\ref{eq1}), where the variables $\gamma_d$ and $\gamma_i$ now represent the received instantaneous power for the desired and interfering links, respectively. Finally, in noise-limited scenarios (i.e. where the interference can be neglected), the OP is given by $OP_{N} (\gamma) = \Pr \left\{ \gamma_d{\rm{ < }}\gamma \right\}$.

\section{System Model II: Wireless Information-Theoretic Security}

\label{Scenario II} 
\subsection{Classical definition}
As in the previous subsection, we consider all the agents to be single antenna transceivers. Using the standing nomenclature in the physical layer security literature \cite{Prabhu2011}, this corresponds to a SISOSE scenario (single-input single-output single eavesdropper). Analogously, we characterize the communication links in Fig. \ref{fig1}b) in terms of their equivalent SNRs, ${\gamma_d=P_T\left|h_d\right|^2/N_0}$ for the \textit{desired} link and ${\gamma_e=P_T\left|h_e\right|^2/N_0}$ for the \textit{eavesdropper} link. Without loss of generality, we assume that the noise power at B and E are equal.

The performance in this scenario is usually characterized by the secrecy capacity $C_s$, defined as
\begin{equation}
C_s  \triangleq C_d  - C_e > 0,  
\end{equation}
\noindent
where $C_d = \log \left( {1 + \gamma_d } \right)$ is the instantaneous capacity of the desired channel and $C_e =  \log \left( {1 + \gamma_e } \right)$ is the instantaneous capacity of the eavesdropper's channel. Secure transmission is achievable provided that ${C_s>0}$ and assuming proper wiretap codes, for which perfect knowledge of both Bob's and Eve's CSI is required at Alice. The probability of strictly positive secrecy capacity is
\begin{align}
\label{eq_Rout_r=0}
P_{s}^{+} &=\Pr \left\{ {C_s  > 0} \right\} = \Pr \left\{ {\log_2 \left( {\frac{{1 + \gamma_d }}{{1 + \gamma_e }}} \right) > 0} \right\} \nonumber \\& = \Pr \left\{ {\gamma_d > \gamma_e} \right\}.
\end{align}
When Eve's CSI is not available at Alice, perfect secrecy cannot always be achieved and secrecy is often analyzed in terms of the secrecy outage probability $P_s$, defined as the probability that communication at a secrecy rate $R_s>0$ cannot be securely achieved \cite{Barros2006,Bloch2008}, i.e.,
\begin{align}
\label{eq_Rout_r}
P_s & = \Pr \left\{ {C_s < R_s } \right\} = \Pr \left\{ {\log_2 \left( {\frac{{1 + \gamma_d }}{{1 + \gamma_e }}} \right) < R_s } \right\} \nonumber \\& = \Pr \left\{ {{{\gamma_d }}{\rm{ < }} (2^{R_s}-1) {{\left(\tfrac{2^{R_s}}{2^{R_s}-1}\gamma_e {\rm{ + }}1\right)}}} \right\}.
 \end{align}
To compute this quantity, Eve's channel statistics (i.e., fading distribution and average SNR) are needed. Whether this information should be available at Alice or not depends on the particular design of the transmission scheme. For instance, Eve's average SNR would be needed at Alice if an adaptive-$R_s$ transmission is implemented so that the outage probability is maintained below a required target. We do not explicitly take this assumption here since other (non-adaptive) schemes could be considered such as, e.g., an offline design of $R_s$ based on worst-case assumptions on Eve's average SNR.

\subsection{Alternative definition}
Although the previous definition of secrecy has been extensively used in the literature, it does not distinguish whether a message transmission is unreliable (i.e., $C_d < R_s$) or not perfectly secure (i.e., $C_s < R_s$). As pointed out in \cite{Zhou2011}, this can be seen from the fact that the event $C_d < R_s$ falls within the secrecy outage event $C_s < R_s$, but the first does not necessarily imply a failure in achieving perfect secrecy. Moreover, if Alice knows that Bob's channel fails to support the secrecy rate, $C_d < R_s$, then transmission should be suspended.

Therefore, an alternative outage formulation is presented in \cite{Zhou2011}, in which the secrecy outage probability is conditioned on the fact that a message is actually being sent, i.e. whenever $\gamma_d$ exceeds some SNR threshold $\mu$:
\begin{align}
\label{eq_PsoAdaptive}
P_{so} & =\Pr \left\{ {C_s <  R_s } | \gamma_d > \mu \right\} \nonumber \\& = \Pr \left\{ {\gamma_d < 2^{R_s} (1+\gamma_e)-1 } | \gamma_d > \mu \right\} \nonumber \\
&= \frac{1}{  \Pr \left\{ \gamma_d > \mu \right\}  } \Pr \left\{ {\mu < \gamma_d < 2^{R_s} (1+\gamma_e)-1 }  \right\}.
\end{align}
\section{Equivalencies between both scenarios}
\label{sec:equiv}
\subsection{Single antenna transceivers}
From the previous expressions, it becomes evident that the secrecy outage probability (with $R_s>0$) defined in (\ref{eq_Rout_r}) is an \textit{equivalent} expression to the one obtained for the wireless system with CCI and BN in (\ref{eq1}); in fact, up to a simple scaling of the underlying random variables, $P_s=OP_{NI}$. For the alternative secrecy formulation in (\ref{eq_PsoAdaptive}), a similar connection can be inferred including some correction terms that are related to the OP without interference, i.e. $OP_N(\gamma)$.

Analogously, we observe that the (complementary) probability of strictly positive secrecy capacity, i.e. $R_s=0$, also follows an \textit{equivalent} expression to the one obtained for the wireless system in interference-limited scenarios, i.e. $P_s^{+}=1-OP_{I}$. Interestingly, we see how doubling the threshold value in the interference setup approximately translates into a 1 bit increase in the secrecy threshold $R_s$. We also see how neglecting the effect of background noise in the interference problem is equivalent to setting the secrecy rate $R_s$ to zero.

The connections among these scenarios are summarized in Table \ref{table1}. Note that we do not take any assumption regarding the distribution of $\gamma_d$, $\gamma_e$ and $\gamma_i$; hence, they can follow arbitrary distributions, and they can also be correlated. A direct implication of the connection unveiled herein is that outage analysis in information-theoretic security problems can be immediately characterized by leveraging the available analytical results for the CCI+BN equivalent problem.

\begin{table*} 
\centering 
\caption{Connection between Scenarios I and II}
\renewcommand{\arraystretch}{1.4}
\begin{tabular}{|l|l|c|}
\hline 
\textsc{Secrecy Metric}&\textsc{Connection CCI+BN}& \textsc{Parameters} \\[0.5ex] 
\hline 
$P_s\triangleq\Pr \left\{ {C_s<R_s }\right\}$& $P_s=OP_{NI}(\gamma)$ & $ \gamma=  2^{R_s} - 1,  \,\,\,\,\,\,\,\,\,  {\gamma_i} =  \frac{{2^{R_s}}}{2^{R_s} - 1} \gamma_e $\\ [4pt] 
\hline 
$P_{so}\triangleq\Pr \left\{ {C_s <  R_s } | \gamma_d > \mu \right\}$& $P_{so}= \frac{OP_{NI}(\gamma) - OP_{N} (\mu)}{1 - OP_{N}(\mu) }$ & $ \gamma=  2^{R_s} - 1,  \,\,\,\,\,\,\,\,\,  {\gamma_i} =  \frac{{2^{R_s}}}{2^{R_s} - 1} \gamma_e$\\[4pt] 
\hline
$P_s^{+}\triangleq\Pr\{C_s>0\}$& $P_s^{+}=1-OP_{I}(\gamma)$ & $\gamma= 1, \,\,\,\,\,\,\,\,\,  {\gamma_i} = \gamma_e$\\ [4pt] 
\hline
\end{tabular}
\label{table1}
\end{table*}

Our approach is thus useful from an analytical point of view: in order to calculate the secrecy outage probability in a new scenario, there is no need to redo the calculations if there is a solution available for the equivalent CCI+BN problem. This is the case for the scenarios investigated in \cite{Ermolova2014}; therefore, a direct application of the secrecy-interference equivalence allows obtaining the secrecy OP in the following cases for which no results are currently available: (i) $\eta$-$\mu$ fading for the A-B link, and $\eta$-$\mu$ fading for the A-E link, (ii) $\eta$-$\mu$ fading for the A-B link, and $\kappa$-$\mu$ fading for the A-E link, (iii) $\kappa$-$\mu$ fading for the A-B link, and $\eta$-$\mu$ fading for the A-E link.  

In addition, the new expressions can be used to optimize the system under secrecy constraints. For example, in \cite{Zhou2011}, for Rayleigh fading channels and under the ``alternative formulation'', the OP expressions were used to design transmission strategies with optimized transmission thresholds and optimized wiretap coding rates. By virtue of the interference-secrecy equivalence, optimized strategies can be designed for more general fading models as mentioned above.

\subsection{Multiple antenna transceivers}
For notational simplicity, we started our analysis with the simplest setup: a scenario where all Alice, Bob, and Eve are equipped with a single antenna. This scenario is referred to as SISOSE (single-input single-output single-eavesdropper) in the literature (see, e.g., \cite{Prabhu2011}). We see that this problem is dual to an interference problem in a SISO setup with a single interferer.

The extension to a SISOME (single-input single-output multiple-eavesdroppers) scenario is direct, since using $L$ antennas at Eve with maximal ratio combining (MRC) reception, which yields ${\gamma_e=\sum_{k=1}^{L} \gamma_{e_k}}$, is equivalent to considering $L$ interfering signals in the dual interference problem \cite{Annamalai2001}, yielding ${\gamma_i=\sum_{k=1}^{L} \gamma_{i_k}}$. 

If we now consider multiple antennas at Bob, then the scenario under analysis is denoted as SIMOME (single-input multiple-output multiple-eavesdroppers). In this case, the SNR at Eve is unaltered and the SNR at Bob is that of a multiantenna receiver, which depends on the implemented combining strategy. It is straightforward to see that the duality also holds in this scenario when having MRC reception: the dual interference problem would be comprised of a receiver with multiple antennas under the effect of multiple interferers. This is a well-investigated problem in the interference-related literature \cite{Cui1999,Aalo2000,Pena2004} when assuming Rayleigh fading channels, whereas the SIMOME counterpart was analyzed in \cite{He2011}.

When considering multiple antennas at Alice and a single antenna at Bob, the scenario is referred to as MISOME (multiple-input single-output multiple-eavesdroppers). In this case, different mechanisms for transmission (e.g. beamforming, transmit antenna selection, space-time block coding) \cite{Khisti2010,Bashar2011,Alves2012} imply different distributions of the effective SNRs and the problem formulation needs some notational changes, but the duality still holds. The dual problem in this case is a MISO communication system affected by multiple interferers\footnote{However, we must note that some of the schemes for multiple antenna transmitters under the presence of eavesdroppers are different that the conventional schemes out of the context of physical layer security; for instance, a masked beamforming technique has been proposed in the context of physical layer security as an improvement over conventional (or naive) beamforming \cite{Khisti2010}.}.

Finally, we consider the more general case where all terminals in the network are equipped with multiple antennas, i.e., MIMOME (multiple-input multiple-output multiple-eavesdroppers). For those schemes that involve rank-1 signaling, i.e., those not involving spatial multiplexing (e.g. beamforming, transmit antenna selection, space-time block coding) \cite{Yang2013,Ferdinand2013}, the channel can be effectively seen as a SISO channel with an underlying SNR distribution possibly more complex. In these cases, the MIMOME problem is equivalent to a MIMO communication system with multiple interferers. A good example can be found in the OP analysis of a MIMO-MRC system affected by interference \cite{Ahn2009}; the OP expressions in \cite{Ahn2009} are a generalization of those for single-antenna transmitters in \cite{Aalo2000}. By virtue of the equivalence unveiled in this paper, the same generalization applies to the MIMOME scenario (in the secrecy context) when a MIMO-MRC configuration is employed in the desired link. 

As we have seen, the interference-eavesdropping connection holds for all the multiple-antenna scenarios described above, i.e., SISOME, SIMOME, MISOME, and MIMOME, provided that rank-1 signaling (without spatial multiplexing) is employed. A generalization to spatial multiplexing schemes is not evident \cite{Lin2014}, and this is indeed an interesting direction for future research.

\section{Application Example: Secrecy analysis in SISOME mixed fading }
Having established the equivalence between these scenarios, we now aim at illustrating how well-known results for cellular systems affected by interference can be translated into analytical results for their wireless information-theoretic counterparts. As an example of application, we investigate the scenario where the fading channel between Alice and Bob is Nakagami-$m$ distributed, whereas the fading experienced by Eve follows an \textit{arbitrary} distribution. Assume that the eavesdropper is equipped with $L_e$ antennas; this is equivalent to consider a set of $L_e$ single-antenna colluding eavesdroppers \cite{Goel2005}, that cooperate to intercept and decode the message.  

The instantaneous SNR $\gamma_{e_k}$ per receive branch ($k=1\ldots L_e$) is modeled as a random variable with mean $\overline{\gamma}_{e_k}$. Furthermore, we assume MRC multichannel reception so that the output SNR $\gamma_e$ is expressed as the sum of the individual per-branch SNRs, i.e. $\gamma_e = \sum\nolimits_{k = 1}^{L_e} \gamma_{e_k}$.

According to Section \ref{sec:equiv}, this problem is equivalent to a wireless system with CCI and BN (System Model I) in which a set of $L_e$ users interfere the desired signal at the receiver B. In this case, the equivalent INR in (\ref{eq1}) is $\gamma_i = \sum\nolimits_{k = 1}^{L_e} \gamma_{i_k}$,
where $\gamma_{i_k}$ represents the SNR of each interfering signal and is (up to a scale factor) directly related to $\gamma_{e_k}$ according to Table \ref{table1}.

Having established this connection, we can now invoke the results in \cite{Annamalai2001} for the OP under CCI and BN, where we just need to replace $\gamma_i$ with the scaled version of $\gamma_e$. For the specific scenario where the desired link undergoes Nakagami-$m$ fading, the OP is directly given in terms of the moment generating function (MGF) of $\gamma_e$, which fortunately can be expressed in closed-form for most common fading distributions. Hence, under the assumptions above, the OP of the secrecy capacity is given by 
\begin{equation}
\label{eq_Ps}
{P_s}  = 1 - {e^{ - p}}\sum\limits_{i = 0}^{m - 1} {\frac{1}{{i!}}\sum\limits_{j = 0}^i {{C_{i,j}}{p^i}} } \frac{{{d^j}}}{{d{p^j}}}{\left. {\left[ {\prod\limits_{k = 1}^{L_e} {{\Phi _k}\left( -p \right)} } \right]} \right|_{p = \tfrac{{m({2^{R_s}} - 1)}}{{{{\bar \gamma }_d}}}}}
\end{equation}
\noindent
where $C_{i,j}={{\left( { - 1} \right)}^j}i!/(j!(i-j)!)$ and $\Phi_k(\cdot)$ represents the MGF of $\gamma_{e_k}$, $k=1,\ldots,L_e$, which are assumed to be independen but can be arbitrarily distributed. Similarly, the probability of strictly positive secrecy capacity is given by 
\begin{equation}
\label{eq_Cs_Generic}
P_s^{+} = \sum\limits_{i = 0}^{m - 1} {\frac{{{(-p)^i}}}{{i!}}} \frac{{{d^i}}}{{d{p^i}}}{\left. {\left[ {\prod\limits_{k = 1}^{L_e} {{\Phi _k}\left( -p \right)} } \right]} \right|_{p = \tfrac{m}{{{{\bar \gamma }_d}}}}}.
\end{equation}
Finally, the secrecy outage probability using the alternative definition in \cite{Zhou2011} can be compactly expressed as
\begin{equation}
\label{eq_Rout}
{P_{so}}  = \frac{1}{1-F_{\gamma_d}(\mu)} \left[ P_s-F_{\gamma_d}(\mu) \right]^+,
\end{equation}
where $\left[ a \right]^+ = \max (0, a)$, $F_{\gamma_d}(\cdot)$ is the cumulative distribution function of $\gamma_d$ (Nakagami-$m$ distributed in this example) and $P_s$ is given by (\ref{eq_Ps}).

Note that the above expressions are given as an example of application, and other scenarios with the desired link arbitrarily distributed can be also analyzed by exploiting the results in \cite{Annamalai2001} along with the equivalences between the interference and secrecy problems. In this case, the analytical results would be given in terms of the MGF of the SNR at Bob. Closed-form expressions for the MGF $\Phi(p)$ and $n$th-order derivative of the MGF $\Phi^{(n)}(p)$ of the signal power in a variety of fading channel models can be easily computed; expressions for the most common fading models are given in Table \ref{table2}.

\begin{table*} 
\centering 
\caption{PDF, MGF and $n$th-order derivative of the MGF of signal power for fading models}
\renewcommand{\arraystretch}{1.4}
\begin{tabular}{|l|l|l|}
\hline 
\textsc{Fading Model}&
 \textsc{MGF, $\Phi(p)$} & \textsc{$n$th-order derivative of the MGF, $\Phi^{(n)}(p)$}\\
\hline
Rayleigh & 
${\left( {1 - p\bar \gamma } \right)^{^{ - 1}}}$
&$\frac{{{{\bar \gamma }^n}n!}}{{{{\left( {1 - p\bar \gamma } \right)}^{n + 1}}}}$

\\ [4pt] 
\hline 
Nakagami-$m$ & 
${ (1-p\bar \gamma /m)^{-m}  }$ 
	&$\frac{{{{\bar \gamma }^n}{m^m}\Gamma \left( {m + n} \right)}}{{{{\left( {m - p\bar \gamma } \right)}^{n + m}}\Gamma \left( m \right)}}$
	 \\[4pt] 

\hline
Rice & 
$\left( \frac{1+K} {1+K-p \bar \gamma} \right) \exp \left(  \frac{pK\bar \gamma} {1+K-p \bar \gamma} \right)$	
& $ \frac{{{{\bar \gamma }^n}{{\left( {n!} \right)}^2}\left( {1 + K} \right)}}{{{{\left( {1 + K - p\bar \gamma } \right)}^{n + 1}}}}\exp \left( {\frac{{pK\bar \gamma }}{{1 + K - p\bar \gamma }}} \right)\sum\limits_{i = 0}^n {\frac{1}{{{{\left( {i!} \right)}^2}\left( {n - i} \right)!}}} {\left( {\frac{{K\left( {1 + K} \right)}}{{1 + K - p\bar \gamma }}} \right)^i}$

 \\ [4pt] 
\hline
\end{tabular}
\label{table2}
\end{table*}

\section{Numerical Results}
\label{numerical}
\begin{figure}[t]
\begin{center}
\includegraphics[width=1\columnwidth]{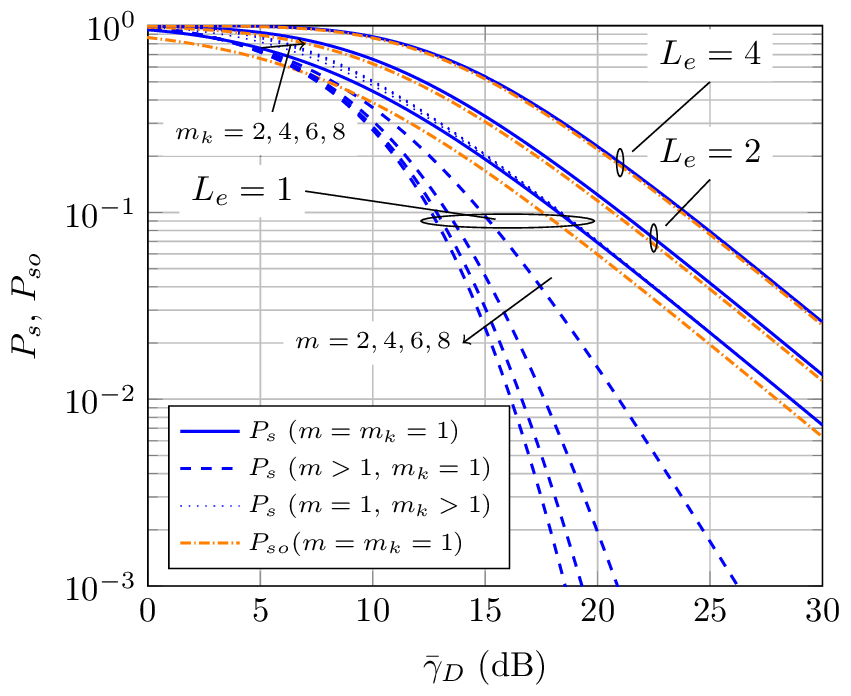}
\caption{OP of secrecy capacity vs $\bar\gamma_d$, for different values of the fading severity index and numbers of receive branches $L_e$. Parameter values are $\overline{\gamma}_{e_k}=5$ dB ($k=1\ldots L_e$) and $R_s=1$.}
\label{fig2}
\end{center}
\end{figure}

We have presented a simple approach for the characterization of the wireless information-theoretic security in terms of the MGF of the SNRs at Eve (or Bob). Next, the derived expressions are evaluated numerically to discuss the main implications that arise in practical scenarios of interest. Let us assume a normalized secrecy rate $R_s=1$ and that Eve supports MRC reception. First, we evaluate the effect of considering a different fading severity for the communication links. The distribution of all links is Nakagami-$m$ with fading severity index $m$ (for the direct link) and $m_k$ (for the $k$th eavesdropper's link). In Fig. \ref{fig2}, the OP of the secrecy capacity derived in (\ref{eq_Ps}) and (\ref{eq_Rout}) are represented as a function of the average SNR at Bob ${\bar \gamma }_{d}$ whereas the average SNR at the eavesdropper links ($\overline{\gamma}_{e_k}$ with $k=1\ldots L_e$) is kept fixed to 5 dB. We also set $\mu=\gamma=2^{R_s}-1$, which is the value of $\mu$ that maximizes the throughput \cite{Zhou2011}.

The number $L_e$ of antennas in the eavesdropper has an important impact on the secrecy capacity outages for both definitions of secrecy; this is expected since a higher number of antennas allows for an increased equivalent SNR after MRC. Therefore, using more antennas at Eve decreases the rate at which secure communication can be established, when the rest of parameters remain fixed; this is in agreement with \cite{Wang2014}. We observe that $P_{so}$ is always lower than the classical definition ($P_{s}$) as the former is conditioned to the fact that a message is being sent ($\gamma_d > \mu$) and it includes the latter for $\mu=0$. The gap between both definitions is reduced as $L_e$ increases since that implies that the effective SNR at Eve (after MRC) gets higher and, consequently, that both $P_{s}$ and $P_{so}$ increase; it is then clear from (\ref{eq_Rout}) that the difference between $P_{so}$ and $P_{s}$ shrinks.

We now aim at analyzing the impact of different fading conditions in the secrecy capacity, which is also shown in Fig. \ref{fig2}. To that end, we vary the fading severity index $m$ for the legitimate communication (between Alice and Bob), while keeping the fading parameter for Eve's link fixed to $m_k=1$, $k=1\ldots L_e$. In the low SNR regime, changing the distribution of the desired link has little effect since in this region the behavior of $P_s$ is dominated by the distribution of $\gamma_{e_k}$. Conversely, in the high SNR regime we see that the down-slope of $P_s$ is proportional to $m$. If we now consider a different fading severity $m_k$ at the eavesdropper link (while keeping $m=1$ for the legitimate communication), we observe a dual effect as compared to the previous case. In the low SNR regime, increasing $m_k$ causes that $\Pr(C_s<R_s)$ grows for a given $\bar\gamma_d$, since the eavesdropper experiences a less severe fading than the legitimate receiver. However, in the high SNR regime the secrecy capacity is practically agnostic to the distribution of $\gamma_{e_k}$. This is explained by the fact that in this region the OP of the secrecy capacity is dominated by the distribution of the desired link, whose average power is much larger than $\bar\gamma_{e_k}$.

\begin{figure}[t]
\begin{center}
\includegraphics[width=1\columnwidth]{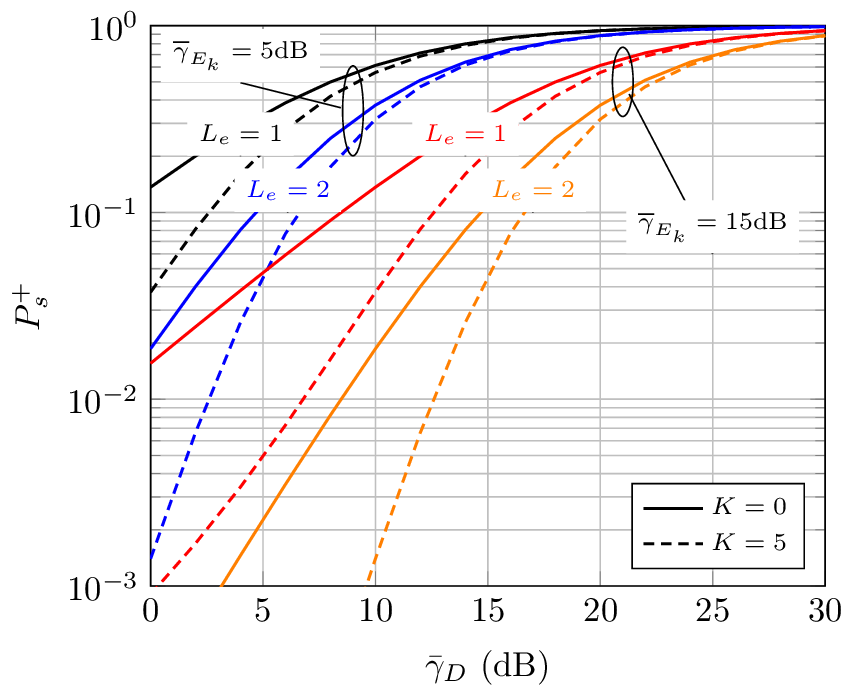}
\caption{Strictly positive secrecy capacity vs $\bar\gamma_d$, for different Rician $K$ values for the eavesdropper link and numbers of receive branches $L_e$; $R_s=1$.}
\label{fig3}
\end{center}
\end{figure}

Fig. \ref{fig3} shows $P_{s}^{+}$, as defined in (\ref{eq_Cs_Generic}), when assuming Rician fading for Eve's link (with different $K$ factors), and Rayleigh fading for the legitimate link. This setup may represent a scenario in which Eve is positioned with a direct line of sight (LoS) to Alice, while Bob is only able to receive the non-LoS component \cite{Prabhu2011}\footnote{This also represents the complementary OP in an interference-limited scenario, i.e. $1-OP_{I}(\gamma)$, with $\gamma_i=\gamma_e$ according to the Table I. In this setting, we are assuming Rician fading for the interferer's link (with different $K$ factors), and Rayleigh fading for the legitimate link. Analogously, this setup may represent a scenario in which the propagation for the signal of interest is non-LoS, whereas the propagation for the interfering signals arriving at the receiver can be modeled as LoS \cite{Annamalai2001}.}. We see that the probability of secure transmission between Alice and Bob is lower as the equivalent average SNR of Eve increases, either after MRC technique (for $L_e>1$) or by increasing the SNR $\overline{\gamma}_{e}$ of its unique channel (for $L_e=1$). Additionally, higher values of the Rician $K$ factor have a negative impact on $P_{s}^{+}$.

\begin{figure}[t]
\begin{center}
\includegraphics[width=1\columnwidth]{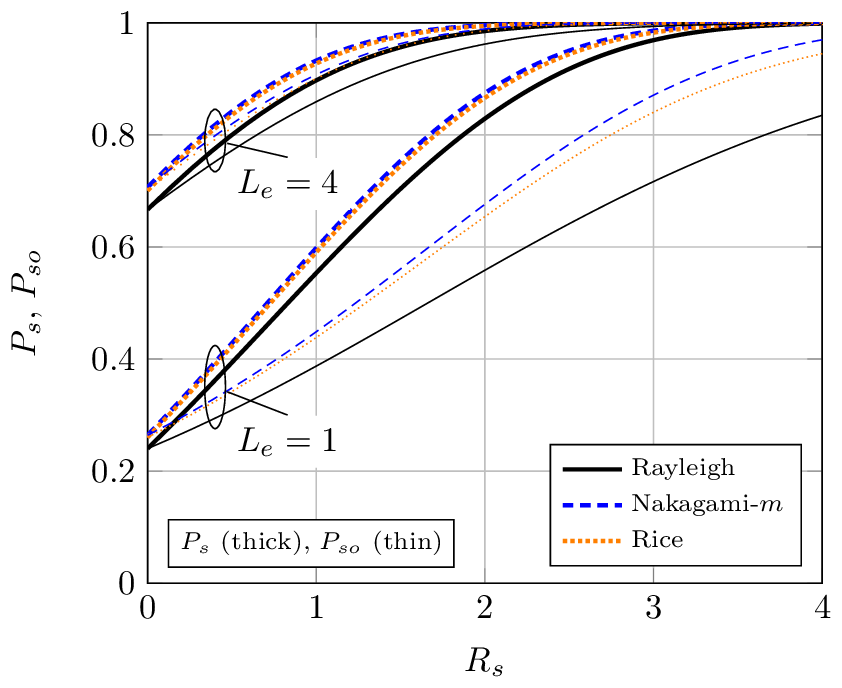}
\caption{Comparison between existing secrecy outage formulations vs $R_s$ for different fading distributions at the eavesdropper link and Rayleigh fading at the desired link. Parameter value $\bar{\gamma}_d=5$ dB, $\bar{\gamma}_e=0$ dB, $m_k=5$, $K=5$.}
\label{fig4}
\end{center}
\end{figure}

Finally, Fig. \ref{fig4} represents the secrecy outage probability as a function of the secrecy rate $R_s$, comparing the original formulation ($P_s$) with the alternative formulation ($P_{so}$) and assuming the same value of $\mu$ as previously. We consider different fading distributions at the eavesdropper link and Rayleigh fading at the desired link. We see from the figure that the outage probabilities predicted by the two formulations differ significantly. To understand such a difference we should recall that $P_{so}$ considers an adaptive transmission scheme based on Bob's CSI.

\section{Conclusion}
\label{Conclusion}
We showed that two relevant yet apparently unrelated problems in information and communication theory are in fact mathematically equivalent. Specifically, we have shown that the problem of calculating the OP with CCI and BN in the presence of $L_e$ interferers is equivalent to the calculation of the OP of the secrecy capacity in the presence of an eavesdropper equipped with $L_e$ antennas.

Similarly, we have also shown that the calculation of the OP in interference-limited scenarios is equivalent to the problem of computing the probability of strictly positive secrecy capacity. Despite their simplicity, these connections have not been previously identified in the literature, and they pave the way towards new results for physical layer security in more general fading models for which the problem of the OP with CCI and BN is well-understood.

The equivalence here unveiled has the potential to be extended to other scenarios in wireless communications, and is an avenue for future research. For instance, the effect of artificial-noise-aided beamforming techniques \cite{Xi2013} in generalized fading scenarios can also be seen as an underlying interference-related problem. Similarly, the coverage probability in wireless cellular networks where the interference is modeled using spatial random models \cite{Andrews2010} could be linked to the information-theoretic characterization in a field of colluding eavesdroppers \cite{Pinto2009}.

\bibliographystyle{ieeetr}
\bibliography{bibfile}

\end{document}